\documentclass[reprint,aps,prl,superscriptaddress,nobibnotes]{revtex4-1}
\usepackage{amsmath}
\usepackage{amssymb}
\usepackage{natbib}
\usepackage{graphicx}
\usepackage{amsfonts}
\usepackage{amsmath}
\setcitestyle{numbers,square}

\begin{document}

\title{Coherent pumping of high momentum magnons by light} 
\author{Fran \v{S}imi\'{c}}
\affiliation{Kavli Institute of NanoScience, Delft University of Technology, 2628 CJ Delft, The Netherlands}  

\author{Sanchar Sharma}
\affiliation{Kavli Institute of NanoScience, Delft University of Technology, 2628 CJ Delft, The Netherlands}  

\author{Yaroslav M. Blanter} 
\affiliation{Kavli Institute of NanoScience, Delft University of Technology, 2628 CJ Delft, The Netherlands}  

\author{Gerrit E. W. Bauer}  
\affiliation{Institute for Materials Research \& WPI-AIMR \& CSRN, Tohoku University, Sendai 980-8577, Japan} 
\affiliation{Kavli Institute of NanoScience, Delft University of Technology, 2628 CJ Delft, The Netherlands}  

\begin{abstract}
We propose to excite a large number of coherent magnons with high momentum in optical cavities. This is achieved by two counterpropagating optical modes that are detuned by the frequency of a selected magnon, similar to stimulated Raman scattering. In sub-mm size yttrium iron garnet spheres, a mW laser input power generates $10^{6}-10^{8}$ coherent magnons. The large magnon population enhances Brillouin light scattering, a probe suitable to access their quantum properties.  
\end{abstract}

\maketitle

Magnets are crucial for fast, non-volatile, and robust data storage as well as candidate materials for logic devices and interconnects \cite{Chumak15}. Magnetic insulators, such as yttrium iron garnet (YIG) \cite{SagaOfYIG}, are interesting since they can transport information over long distances via spin waves quantized into magnons \cite{Kajiwara10,Cornelissen15}, without the Ohmic dissipation of spin transport in metals. The magnons couple to microwaves \cite{SoykalPRL10,Zhang14,TabuchiHybrid14}, electric currents \cite{Kajiwara10,Chumak15,Bai15}, mechanical motion \cite{Kittel58,Akash15,Kikkawa16,Zhang16}, and light \cite{Borovik82,HansteenMO}. The high crystal quality of YIG promises long coherence times \cite{Zhang14,TabuchiHybrid14}, opening prospects for `quantum magnonics' \cite{Tabuchi_QMag}, the field that strives to employ magnons to store, process, and transfer information in a quantum coherent manner. Photons can become a coherent interface to manipulate and probe these magnons.  

The GHz magnons in ferro(ferri)magnets interact with light by inelastic (Brillouin) light scattering (BLS) \cite{Borovik82}. By selecting the wave vector of the input and output photons, e.g. by an optical cavity, specific magnons modes can be excited \cite{WGMOptoMag}. The interaction can be large enough \cite{Jasmin_Vortex,OptimalMatching} to cool \cite{OMagCool} or herald (generating single magnon states) \cite{Viktor_Heralding} them, making BLS a promising probe into their quantum nature. Present experiment focus on the long wavelength `Walker' (including the `Kittel') magnons in optical resonators \cite{Osada_OMag,Zhang_OMag,James_OMag,Osada_LowL,James_LowL}. These have a small overlap with the light fields and corresponding low intrinsic scattering efficiency, but become observable because a large magnon density can be resonantly excited by microwaves. On the other hand, magnons with wavelengths $\sim100 - 500\,$nm in the dipolar-exchange regime have almost perfect overlap with the photon modes in magnetic spheres \cite{OptimalMatching}, but couple only very weakly to microwaves (as do the relevant magnons in magnetic vortices \cite{Jasmin_Vortex}).  

\begin{figure}[ptb]
\begin{equation*}
\includegraphics[width=\columnwidth]{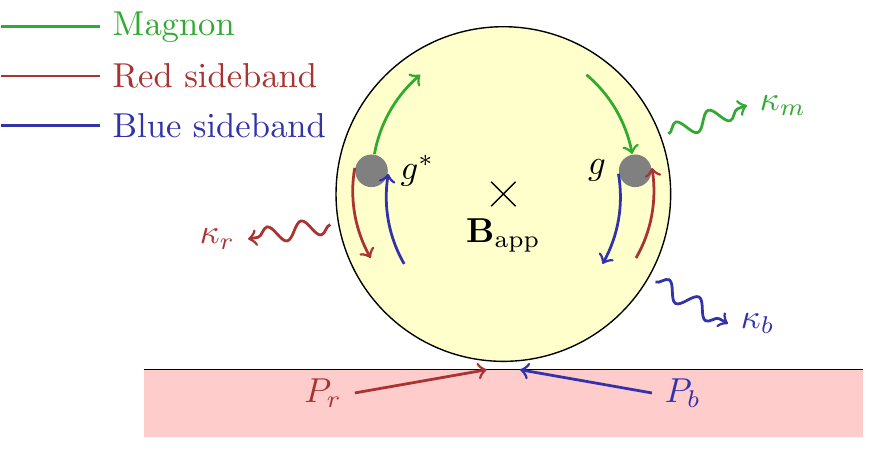}
\end{equation*}
\caption{A (massive) sphere of a magnetic insulator, such as YIG, with a proximity optical coupler, such as a fiber or a prism. Two oppositely propagating laser beams excite two whispering gallery modes with decay rates $\kappa_{r,b}$. The photon-magnon scattering coherently amplifies the magnon amplitude competing with the thermalization rate $\kappa_{m}$.}
\label{Fig:Sphere}
\end{figure}

Here, we propose to coherently pump a large number ($\sim10^6 - 10^8$) of high-momentum magnons by optical lasers, similar to the resonant excitation of Kittel magnons by microwaves. We exploit the torques exerted by light on the magnetization by the inverse Faraday and Cotton-Mouton effects \cite{IMO_Rev}, which are proportional to the intensity of the electric field component \cite{IMO_Rev} or, more precisely, the product of the photon numbers at the incident and scattered frequencies. Exposing the sample to two phase-coherent lasers that differ in frequency by a magnon excitation strongly enhances Brillouin scattering \cite{Richard_17}. Here we develop the theory of stimulated light scattering by magnons in optical resonators such as sketched in Fig.~\ref{Fig:Sphere}. Two counter-propagating lasers feed whispering gallery modes (WGMs) of a YIG sphere via a proximity coupler such as a fiber or a prism \cite{James_YIGWGM,Osada_OMag,Zhang_OMag,James_OMag}. The WGMs are separated spectrally by $\sim1 - 10\,$GHz, which can be easily tuned into resonance with a magnon by an applied magnetic field. The two populated WGMs form a spatially-periodic torque field that excites magnons with matching wavelength. While we focus here on spherical magnets, the formalism is valid for any magneto-optical cavity, including planar \cite{Tianyu_Planar,PAP_OMagCav} and cylindrical \cite{Jasmin_Vortex} geometries.  

We consider a minimal model of two WGM modes $\{W_r,W_b\}$ resonantly interacting with a single magnon mode $M$ [see Fig.~\ref{Fig:Sphere}]. We first formulate heuristic rate equations for the magnon number, $n_m^{\mathrm{(sc)}}$ (`sc' stands for semi-classical), followed by a more rigorous quantum Langevin treatment. In the steady state, the energy balance of the processes in Fig.~\ref{Fig:Sphere} leads to the photon number in the blue sideband $W_b$ with frequency $\omega_b$ \cite{COMech_Rev} 
\begin{equation}
	 N_b = \frac{4K_b}{(\kappa_b + K_b)^2}\frac{P_b}{\hbar\omega_b},\label{nx:Def} 
\end{equation} 
which is governed by the input light power $P_b$, the decay rate $\kappa_b$ in the isolated sphere, and the leakage rate $K_b$ into the proximity coupler. An analogous expression holds for the photon number $N_r$ in the red sideband $W_r$. Since optomagnonic couplings are small, we disregarded the backaction exerted by magnons on photons. The reaction rate for anti-Stokes scattering $W_r + M\rightarrow W_b$ is $R_b = R_b^{(0)}n_m^{\mathrm{(sc)}}N_r(N_b + 1),$ while for the reverse (Stokes) scattering $R_r = R_r^{(0)}(n_m^{\mathrm{(sc)}} + 1)(N_r + 1)N_b$. According to the Fermi's golden rule, $R_{b,r}^{(0)} = 2\pi\left\vert g\right\vert^2 \Lambda_{b,r}(\Delta)$, where $g$ is the matrix element of the Hamiltonian between initial and final states (see below), the detuning $\Delta\equiv\omega_b - \omega_r - \omega_m$, and 
\begin{equation}
	 \Lambda_{b,r} = \frac{1}{2\pi}\frac{\left( \kappa_{b,r} + K_{b,r}\right)}{\Delta^2 + (\kappa_{b,r} + K_{b,r})^2/4},\label{Def:Rc} 
\end{equation} 
with $\left( \kappa_{b,r} + K_{b,r}\right)^{- 1}$ as the photon's lifetime.  

Magnons are lost at a rate $R_{\mathrm{eq}} = \kappa_m\left( n_m^{\mathrm{(sc)}} - n_{\mathrm{eq}}\right) $ where $\kappa_m^{- 1}$ is the magnon lifetime and $n_{\mathrm{eq}}$ is the equilibrium (Planck) distribution 
\begin{equation}
	 n_{\mathrm{eq}} = \left[ \exp\left( \frac{\hbar\omega_m}{k_BT}\right) - 1\right]^{- 1} . \label{Def:nmth} 
\end{equation} 
In the steady state $R_b + R_{\mathrm{eq}} = R_r$ 
\begin{equation}
	 n_m^{\mathrm{(sc)}} = \frac{R_r^{(0)}N_b\left( N_r + 1\right) + \kappa_mn_{\mathrm{eq}}}{\kappa_m + R_b^{(0)}N_r\left( N_b + 1\right) - R_r^{(0)}\left( N_r + 1\right) N_b} . \label{Est} 
\end{equation} 
Eq.~(\ref{Est}) agrees with the more rigorous result below only when $R_b^{(0)} = R_r^{(0)}$, because here we ignored the correlation between the forward and backward reactions. Furthermore, the above treatment does not distinguish between coherent and thermal magnons. For sufficiently large $N_{r,b}$, $n_m^{\mathrm{(sc)}}$ may diverge which is an artifact of ignoring magnon non-linearities, but such large drives are unrealistic (shown below).  

We consider the 3-particle Hamiltonian $\hat{H} = \hat{H}_0 + \hat{H}_{\mathrm{om}}$ with non-interacting part 
\begin{equation}
	 \hat{H}_0 = \hbar\omega_r\hat{a}_r^{\dagger}\hat{a}_r + \hbar\omega_b\hat{a}_b^{\dagger}\hat{a}_b + \hbar\omega_m\hat{m}^{\dagger}\hat{m}, 
\end{equation} 
where $\{\hat{a}_r,\hat{a}_b,\hat{m}\}$ are the annihilation operators for $\{W_r,W_b,M\},$ respectively. To leading order in the magnon operators the optomagnonic Hamiltonian is \cite{WGMOptoMag} 
\begin{equation}
	 \hat{H}_{\mathrm{om}} = \hbar g\hat{a}_r\hat{a}_b^{\dagger}\hat{m} + \hbar g^{\ast}\hat{a}_r^{\dagger}\hat{a}_b\hat{m}^{\dagger} . 
\end{equation} 
In the Heisenberg picture, the statistical averages $\left\langle \hat{X}(t)\right\rangle = \mathrm{Tr}\left[ \hat{X}(t)\hat{\rho}\right] $, where the density matrix $\hat{\rho} = \hat{\rho}_0$ is a direct product of an arbitrary state of the sphere (magnons and WGMs) and a coherent photon state of the laser input.  

The equation of motion for the blue sideband envelope operator $\hat{W}_b\overset{\triangle}{=}\hat{a}_be^{i\omega_bt}$ reads \cite{Florian07,OMagCool,Viktor_Heralding} 
\begin{equation}
	 \frac{d\hat{W}_b}{dt} =  - ig\hat{W}_r\hat{M}e^{i\Delta t} - \frac{\kappa_b + K_b}{2}\hat{W}_b - \sqrt{\kappa_b}\hat{b}_b + \sqrt{K_b}\hat{A}_b,\label{EOM:Wb} 
\end{equation} 
where the first term on the r.h.s. is the optomagnonic scattering generated by the commutator $\left[ \hat{a}_b,\hat{H}\right] $ in the Heisenberg equation. The second term is the decay of photons inside the sphere, $\propto\kappa_b$, and into the coupler, $\propto K_b$. $\hat{b}_b$ is the annihilation field operator of a bath mode that interacts with $W_b$ satisfying the commutation relations $\left[ \hat{b}_b(t),\hat{b}_b^{\dagger}(t^{\prime})\right] = \delta(t - t^{\prime})$ and averages $\left\langle \hat{b}_b(t)\right\rangle = \left\langle \hat{b}_b^{\dagger}(t^{\prime})\hat{b}_b(t)\right\rangle = 0$. Without input $K_b = 0$ and optomagnonic coupling $g = 0$, the steady state of Eq.~(\ref{EOM:Wb}) is the thermal equilibrium state \cite{GardinerOrig} with no photons since $k_B T\ll\hbar\omega_b$. The input field operator $\hat{A}_b$ of the propagating photons in the coupler \cite{Florian07,OMagCool,Viktor_Heralding} satisfies the commutation relations $\left[ \hat{A}_b(t),\hat{A}_b^{\dagger}(t^{\prime})\right] = \delta(t - t^{\prime})$, with average 
\begin{equation}
	 \left\langle \hat{A}_b(t)\right\rangle = \sqrt{\frac{P_b}{\hbar\omega_b}}e^{i\mathcal{W}_b(t)}, 
\end{equation} 
and correlator 
\begin{equation}
	 \left\langle \hat{A}_b^{\dagger}(t^{\prime})\hat{A}_b(t)\right\rangle = \frac{P_b}{\hbar\omega_b}e^{i(\mathcal{W}_b(t) - \mathcal{W}_b(t^{\prime}))} . 
\end{equation} 
The photons suffer from phase noise that we model by a classical random walk $\mathcal{W}_b(t) = \sqrt{\kappa_{\mathrm{ph}}}\int_0^t\mathcal{N} (x)dx,$ with dephasing rate $\sqrt{\kappa_{\mathrm{ph}}}$. $\kappa_{\mathrm{ph}} /(2\pi)$ typically ranges from Hz to MHz \cite{Tur_PhaseNoise}, much smaller than the typical inverse lifetimes in a resonator $\kappa_{\mathrm{ph}} \ll\kappa_b\sim2\pi\times0 . 1 - 1\,$GHz. The phase noise is taken to be white with $\left\langle \mathcal{N}\right\rangle_{\mathrm{cl}} = 0$ and $\left\langle \mathcal{N}(t)\mathcal{N}(t^{\prime})\right\rangle_{\mathrm{cl}} = \delta(t - t^{\prime})$.  

Since Eq.~(\ref{EOM:Wb}) is linear, $\hat{W}_b(t) = \hat{W}_{b,\mathrm{opt}}(t) + \hat{W}_{b,\mathrm{om}}(t)$, with optical contribution at large times $t\gg1/\kappa_b$ being, 
\begin{align}
	 \hat{W}_{b,\mathrm{opt}}(t) &=  - \int_0^te^{- (\kappa_b + K_b)(t - \tau)/2} \nonumber \\
	 &\mathbin{\phantom{=}} \left[ \sqrt{\kappa_b}\ \hat{b}_b(\tau) + \sqrt{K_b}\ \hat{A}_b (\tau)\right] d\tau \label{Wbopt} 
\end{align} 
includes the thermal noise and input from the coupler. In the steady state and for $\kappa_{\mathrm{ph}}\ll\kappa_b$, we get the commutation relations 
\begin{equation}
	 \left[ \hat{W}_{b,\mathrm{opt}}\left( t\right) ,\hat{W}_{b,\mathrm{opt}}^{\dagger}\left( t^{\prime}\right) \right] = e^{- (\kappa_b + K_b )\left\vert t - t^{\prime}\right\vert /2}, 
\end{equation} 
the average 
\begin{equation}
	 \left\langle \hat{W}_{b,\mathrm{opt}}(t)\right\rangle = \sqrt{N_b}e^{i\mathcal{W}_b(t)}, 
\end{equation} 
and correlator 
\begin{equation}
	 \left\langle \hat{W}_{b,\mathrm{opt}}^{\dagger}(t^{\prime})\hat{W}_{b,\mathrm{opt}}(t)\right\rangle = N_be^{i(\mathcal{W}_b(t) - \mathcal{W}_b(t^{\prime}))} 
\end{equation} 
with $N_b$ from Eq.~(\ref{nx:Def}). The optomagnonic scattering $W_r + M\rightarrow W_b$ contributes 
\begin{equation}
	 \hat{W}_{b,\mathrm{om}}(t) =  - ig\int_0^te^{- (\kappa_b + K_b)(t - \tau)/2}\hat{W}_r(\tau)\hat{M}(\tau)e^{i\Delta\tau}d\tau . \label{Wbom} 
\end{equation} 
For the red sideband $\hat{W}_r(t) = \hat{W}_{r,\mathrm{opt}}(t) + \hat{W}_{r,\mathrm{om}}(t),$ with $\hat{W}_{r,\mathrm{opt}}(t)$ analogous to Eq. (\ref{Wbopt}) and scattering contribution 
\begin{equation}
	 \hat{W}_{r,\mathrm{om}}(t) =  - ig^{\ast}\int_0^te^{- (\kappa_r + K_r )(t - \tau)/2}\hat{W}_b(\tau)\hat{M}^{\dagger}(\tau)e^{- i\Delta\tau} d\tau . \label{Wrom} 
\end{equation}

The magnon envelope operator $\hat{M}(t)\overset{\triangle}{=}\hat{m}(t)e^{i\omega_mt}$ obeys 
\begin{equation}
	 \frac{d\hat{M}}{dt} =  - ig^{\ast}\hat{W}_r^{\dagger}\hat{W}_be^{- i\Delta t} - \frac{\kappa_m}{2}\hat{M} - \sqrt{\kappa_m}\hat{b}_m ,\label{FirstEOM} 
\end{equation} 
where the stochastic magnetic field, $\hat{b}_m(t)$, is generated by magnon-phonon \cite{Simon_Damping}, magnon-magnon \cite{Schlomann58,Kasuya_LeCraw}, surface roughness \cite{SurRough} and (rare earth) impurity scattering \cite{Sparks_FMR,Boventer18,Maier_Linewidth,Kosen19}. When $k_BT/\hbar \gg\kappa_m$, which for $\kappa_m\sim2\pi\times1\,$MHz \cite{Osada_OMag,Zhang_OMag,James_OMag} means $T\gg50\,\mathrm{\mu}$K, we can write $\left\langle \hat{b}_m(t)\right\rangle = 0$, $\left\langle \hat{b}_m^{\dagger}(t^{\prime})\hat{b}_m(t)\right\rangle = n_{\mathrm{eq}} \delta(t - t^{\prime})$ and $\left\langle \hat{b}_m(t^{\prime})\hat{b}_m^{\dagger}(t)\right\rangle = (n_{\mathrm{eq}} + 1)\delta(t - t^{\prime})$, with average magnon number $n_{\mathrm{eq}}$ [see Eq.~(\ref{Def:nmth})]. When $g = 0$, the steady state of Eq.~(\ref{FirstEOM}) is the Planck distribution of the magnon number at temperature $T$ \cite{GardinerOrig}, given in Eq.~(\ref{Def:nmth}).  

The optical torque $\propto g^{\ast}$ in Eq.~(\ref{FirstEOM}) generates coherent magnons. To leading order in $g/\kappa_{r,b}$, 
\begin{equation}
	 \frac{d\left\langle \hat{M}\right\rangle}{dt} =  - i\bar{\omega}\left\langle \hat{M}\right\rangle - ig^{\ast}\sqrt{N_rN_b}e^{- i\Delta t + i\mathcal{W} (t)} - \frac{\kappa_{\mathrm{eff}}}{2}\left\langle \hat{M}\right\rangle ,\label{avgM:EOM} 
\end{equation} 
where 
\begin{equation}
	 \bar{\omega} = \left\vert g\right\vert^2\Delta\left( \frac{4N_b}{4\Delta^2 + \left( \kappa_r + K_r\right)^2} - \frac{4N_r}{4\Delta^2 + \left( \kappa_b + K_b\right)^2}\right) , 
\end{equation} 
is a shift in the magnon frequency, $\mathcal{W} = \mathcal{W}_b - \mathcal{W}_r$ is the phase noise with variance $2\kappa_{\mathrm{ph}}$, and the effective damping $\kappa_{\mathrm{eff}} = \kappa_m + \bar{\kappa}_b - \bar{\kappa}_r . $ Here 
\begin{equation}
	 \bar{\kappa}_b = \frac{4\left\vert g\right\vert^2N_r\left( \kappa_b + K_b\right)}{4\Delta^2 + \left( \kappa_b + K_b\right)^2}\label{DampingMag} 
\end{equation} 
is proportional to the reaction rate of $W_r + M\rightarrow W_b$ [see Eq.~(\ref{Def:Rc})] and $\bar{\kappa}_r$ is given by $r\leftrightarrow b$. Eq.~(\ref{avgM:EOM}) leads to the steady state 
\begin{equation}
	 \lim_{t\rightarrow\infty}\left\langle \hat{M}(t)\right\rangle = \frac{- ig^{\ast}\sqrt{N_rN_b}}{i(\bar{\omega} - \Delta) + \kappa_{\mathrm{ph}} + \kappa_{\mathrm{eff}}/2}e^{- i\Delta t + i\mathcal{W}(t)},\label{CohM:res} 
\end{equation} 
where we assumed ergodicity of $\mathcal{W}$. The phase noise of the input laser fields is imprinted on the magnon amplitude.  

We estimate the magnitude of the effects for an input laser with typical vacuum wavelength $\sim1\,\mathrm{\mu}$m and $\omega_r \approx \omega_b \approx \omega_{\mathrm{opt}} = 2\pi\times300\,$THz. For a YIG sphere, the optical quality can be as high as $\omega_r/\kappa_r = \omega_b/\kappa_b = 10^6$ \cite{Zhang16} and is limited by light absorption (for frequencies at which the magneto-optical coupling is significant). The magnon linewidth $\kappa_m = 2\pi\times1\,$MHz and we adopt the optomagnonic coupling $\left\vert g\right\vert = 2\pi\times200\,$Hz \cite{OptimalMatching} for a sphere of radius $R = 300\,\mathrm{\mu}$m (with $\left\vert g\right\vert \propto1/R$). We assume low phase noise $\kappa_{\mathrm{ph}}\ll\kappa_m$ which can otherwise be absorbed into $\kappa_{\mathrm{eff}}$, cf. Eq.~(\ref{CohM:res}). An external magnetic field can tune $\omega_m$ into resonance at $\Delta = 0$. For impedance-matched optical coupling $\kappa_{r,b} = K_{r,b} = \kappa_{\mathrm{opt}}$, the total magnetic damping  
\begin{equation}
	 \kappa_{\mathrm{eff}} = \kappa_m\left( 1 + \frac{P_r - P_b}{P_{\mathrm{sat}}}\right) 
\end{equation} 
with saturation power (to be interpreted below) 
\begin{equation}
	 P_{\mathrm{sat}} = \frac{\hbar\kappa_m\omega_{\mathrm{opt}}\kappa_{\mathrm{opt}}^2}{2\left\vert g\right\vert^2} = 1\,\mathrm{W} . 
\end{equation} 
For moderate $P_{r,b}\sim1 - 10\,$mW, $\kappa_{\mathrm{eff}} \approx \kappa_m$ is limited by the intrinsic (Gilbert) damping of the magnet. For the large coupling $\left\vert g\right\vert = 2\pi\times4\,$kHz predicted for a magnetic vortex in a thin magnetic disk \cite{Jasmin_Vortex}, $P_{\mathrm{sat}} = 3 . 5\,$mW.  

Our main result is the number of coherently excited magnons 
\begin{equation}
	 n_{\mathrm{c}} = \lim_{t\rightarrow\infty}\left\vert \left\langle \hat{M}(t)\right\rangle \right\vert^2 = \frac{P_rP_b}{P_{\mathrm{crit}}^2},\label{nc:res} 
\end{equation} 
in terms of the critical power 
\begin{equation}
	 P_{\mathrm{crit}} = \frac{\hbar\kappa_{\mathrm{eff}}\omega_{\mathrm{opt}} \kappa_{\mathrm{opt}}}{2\left\vert g\right\vert},\label{Pcrit} 
\end{equation} 
which is a measure for the input power required to generate significant coherent dyamics. It is smaller than $P_{\mathrm{sat}}$ by a factor $\kappa_{\mathrm{opt}}/\left\vert g\right\vert \sim10^6$. With $\kappa_{\mathrm{eff}} \approx \kappa_m$, $P_{\mathrm{crit}} = 1\,\mathrm{\mu}$W is in experimental reach. We predict a large $n_c = 10^6 - 10^8$ for $P_{r,b} \sim1 - 10\,\mathrm{mW}$. In a magnetic vortex \cite{Jasmin_Vortex}, $P_{\mathrm{crit}} = 50\,$nW and $n_c = 5\times(10^8 - 10^{10})$.  

Next we demonstrate that the coherence of the excited magnons is very high (in the absence of absorption heating by the lasers), i.e. the fluctuations around the coherent component $\delta\hat{M} = \hat{M} - \langle \hat{M} \rangle$ are very small, by solving Eq.~(\ref{FirstEOM}). We employ a weak coupling approximation \cite{OMagCool} by expanding up to the leading terms in $\hat{W}_{x,\mathrm{om}}$. When $\delta\hat{M}$ varies much slower than $\kappa_{r,b}$ (shown a posteriori to be equivalent to high optical damping $\kappa_{r,b}\gg\kappa_{\mathrm{eff}}$) we can replace $\delta\hat{M}(\tau)\rightarrow\delta\hat{M}(t)$ in the expression of photons Eqs.~(\ref{Wbom},\ref{Wrom}). Furthermore, we ignore correlations between photons and magnons beyond second order in $g$, which is equivalent to replacing photon operators by their mean-field average (see \cite{OMagCool} for intermediate steps). Then Eq.~(\ref{FirstEOM}) reduces to 
\begin{equation}
	 \frac{d}{dt}\delta\hat{M} =  - \left( i\bar{\omega} + \frac{\kappa_{\mathrm{eff}}}{2}\right) \delta\hat{M} - \sqrt{\kappa_{\mathrm{eff}}}\hat{b}_{\mathrm{eff}}\label{deltaM:EOM} 
\end{equation} 
where $\bar{\omega}$ and $\kappa_{\mathrm{eff}}$ are defined below Eq.~(\ref{avgM:EOM}) and the cumulative noise 
\begin{multline}
	 \sqrt{\kappa_{\mathrm{eff}}}\hat{b}_{\mathrm{eff}}(t) = \sqrt{\kappa_m}\hat{b}_m(t) \nonumber \\
	 + ig^{\ast}e^{- i\Delta t}\left( \hat{W}_{r,\mathrm{opt}}^{\dagger} (t)\hat{W}_{b,\mathrm{opt}}(t) - \sqrt{N_bN_r}e^{i\mathcal{W}(t)}\right) . 
\end{multline} 
The statistics for $\kappa_{r,b}\gg\kappa_{\mathrm{eff}}$: $\left\langle \hat{b}_{\mathrm{eff}}\right\rangle = 0$, $\left\langle \hat{b}_{\mathrm{eff}}^{\dagger}(t^{\prime})\hat{b}_{\mathrm{eff}}(t)\right\rangle \approx n_{\mathrm{th}}\delta(t - t^{\prime})$, and $\left\langle \hat{b}_{\mathrm{eff}}(t)\hat{b}_{\mathrm{eff}}^{\dagger}(t^{\prime})\right\rangle \approx \left( n_{\mathrm{th}} + 1\right) \delta(t - t^{\prime})$, 
\begin{equation}
	 n_{\mathrm{th}} = \frac{\kappa_mn_{\mathrm{eq}} + \bar{\kappa}_r}{\kappa_{\mathrm{eff}}}\rightarrow\frac{n_{\mathrm{eq}} + P_b/P_{\mathrm{sat}}}{1 + (P_r - P_b)/P_{\mathrm{sat}}},\label{Thermal} 
\end{equation} 
and $\rightarrow$ holds for impedance-matched optical coupling $\kappa_{r,b} = K_{r,b}$. Eq.~(\ref{deltaM:EOM}) is equivalent to the equation of motion for magnons in equilibrium [Eq.~(\ref{FirstEOM}) with $g = 0$] after substituting $\omega_m\rightarrow\omega_m + \bar{\omega}$, $\kappa_m\rightarrow\kappa_{\mathrm{eff}}$, and $\hat{b}_m\rightarrow\hat{b}_{\mathrm{eff}}$. Therefore in the steady state 
\begin{equation}
	 \lim_{t\rightarrow\infty}\left\langle \delta\hat{M}^{\dagger}(t)\delta\hat{M}(t)\right\rangle = n_{\mathrm{th}}, 
\end{equation} 
justifying the notation $n_{\mathrm{th}}$. At $P_b - P_r = P_{\mathrm{sat}}$, the magnon damping $\kappa_{\mathrm{eff}}$ vanishes and the magnon number $n_{\mathrm{th}}$ diverges.  The system becomes unstable and magnon non-linearities should be taken into account \cite{Silvia_OMag}. For $T\sim1\,$K, $n_{\mathrm{eq}}\sim10$ and $n_{\mathrm{th}}\sim n_{\mathrm{eq}}$ for realistic powers $P_{r,b}\ll P_{\mathrm{sat}}$. Thus, $n_{\mathrm{th}}\ll n_c$, i.e. the coherently precessing magnetization is accompanied only by a small thermal cloud.  

A large magnon population increases the BLS scattering cross section \cite{Osada_OMag,Zhang_OMag,James_OMag}: the uniform mode can be observed in BLS by exciting $>10^{12}$ magnons by microwaves \cite{Zhang_OMag} in spite of the small optomagnonic coupling $g<2\pi\times5\,$Hz. We consider now the enhancement of BLS by the high-momentum mode $M$ that is coherently excited as discussed above. This can be measured by a third (probe) beam that couples to another optical WGM. Typically, only one of the sidepeaks dominates \cite{WGMOptoMag}, with a ratio of scattered to incident (impedance-matched) photons 
\begin{equation}
	 S = \frac{\left\vert g^{\prime}\right\vert^2\left( n_c + n_{\mathrm{th}}\right)}{\kappa_{\mathrm{opt}}^2}, 
\end{equation} 
where $g^{\prime}$ is the coupling of the probe WGMs with the $M$-magnons and $\kappa_{\mathrm{opt}}$ is a typical optical linewidth. For $g^{\prime} = 2\pi\times200\,$Hz we require $P_{r,b} = 5\,$mW for a signal that exceeds the noise background $S_{\mathrm{noise}}\sim10^{- 5}$ \cite{Zhang_OMag} . A threefold larger $\{g,g^{\prime}\}$ when reducing the radius to $100\,\mathrm{\mu}$m increases $S$ by two orders of magnitude (because $n_c\propto|g|^2$). For thin magnetic\ disks with $\left\vert g\right\vert = 2\pi\times4\,$kHz \cite{Jasmin_Vortex} $S\sim1$.  

Coherent magnons can also be excited by femtosecond laser pulses with a frequency spectrum that overlaps with the two WGMs, a process known as \textquotedblleft impulsive stimulated Raman scattering\textquotedblright \ \cite{Hovhannisyan03,HansteenMO,IMO_Rev}. Time periodic and phase-coherent laser pulses (frequency combs) \cite{Theodor_Nobel,KippenbergCombs} have a spectrum of sharp and periodic peaks whose period can be tuned to a magnon frequency. These techniques can achieve high laser intensities, but are less selective.  

In summary, we show that two counter-propagating slightly detuned lasers can excite a large $\sim10^6 - 10^8$ number of coherent magnons with sub-$\mathrm{\mu}$m wavelengths in a conventional experimental setup of a proximity-coupled YIG sphere of radius $\sim300\,\mathrm{\mu}$m. The consequent enhancement of the BLS cross section makes it experimentally feasible to observe. The coherent optical excitation of short-wavelength magnons with high group velocities can serve as an improved interface between light and spintronic devices in quantum domain.  

%


\end{document}